\documentclass{JHEP3} 
 
\usepackage[utf8]{inputenc} 
\usepackage[T1]{fontenc} 
\usepackage{slashed} 

\usepackage{amssymb,amsfonts,amsmath,amsthm} 
 
\usepackage{slashed} 
 
\usepackage{graphicx} 

\preprint{DO-TH 10/04 \\ TTK-10-20}

\title{How large can the SM contribution to CP violation in $D^0-\bar D^0$ mixing be?} 
 
\author{M. {Bobrowski}$^a$, A. Lenz$^{b,a}$, J. Riedl$^a$ and J. {Rohrwild}$^{a,c}$\\ 
$^a$ Institut f\"ur Theoretische Physik, Universit\"at Regensburg, \\ D-93040 Regensburg, Germany\\
$^b$ Institut f\"ur Physik,   Technische Universit\"at Dortmund,   \\ D-42211 Dortmund, Germany\\
$^c$ Institut f\"ur Theoretische Teilchenphysik und Kosmologie, RWTH Aachen University,
     \\ D-52056 Aachen, Germany}

\abstract{ 
We investigate the maximum size of CP violating effects in $D$-mixing
within the Standard Model (SM), using Heavy Quark Expansion (HQE) as
theoretical working tool.
For this purpose we determine the leading HQE contributions and also
$\alpha_s$ corrections as well as subleading $1/m_c$ corrections to the
absorptive part of the mixing amplitude of neutral $D$ mesons.
It turns out that these contributions to $\Gamma_{12}$ do not vanish in
the exact SU$(3)_\mathrm{F}$ limit. Moreover, while the leading HQE terms give a
result for $\Gamma_{12}$ orders of magnitude lower than the current
experimental value, we do find a sizeable phase.
In the literature it was suggested that higher order terms in the HQE
might be much less affected by the severe GIM cancellations of the
leading terms; it is even not excluded that these higher order terms can
reproduce the experimental value of $y$. If such an enhancement
is realized in nature, the phase discovered in the leading HQE terms can have a
sizeable effect. 
Therefore, we think that statements like: {\it "CP violating effects in
$D$-mixing of the order of $10^{-3}$ to $10^{-2}$ are an unambigous sign
of new physics"}---given our limited knowlegde of the SM prediction---are 
premature.
Finally, we give an example of a new physics model that can
enhance the leading HQE terms to $\Gamma_{12}$ by one to two orders of
magnitude.
}

\begin{document} 
%
 
\section{Introduction} 
 
Meson-antimeson oscillations have long since provided a 
rich area for theoretical studies.   
In an application of pure quantum mechanics, the concept of neutral  
kaon mixing \cite{GellMann:1955jx} led to the understanding  
of the observed CP violation in the decay  
$K_{\rm L} \to \pi \pi$ \cite{Christenson:1964fg}.  
Furthermore, hints to the mass of the $c$ quark 
\cite{Gaillard:1974hs} 
were obtained from $K$ mixing before the first evidence of the $J/\Psi$  
\cite{Aubert:1974js, Augustin:1974xw}. 
Also $B_d-\bar B_{d}$ mixing \cite{Albrecht:1987dr, Albajar:1986it}   
provided information on the mass of the $t$ quark prior to its 
discovery \cite{Abachi:1995iq, Abe:1995hr}. 
The latter was possible due to the sensitivity of  
meson-antimeson mixing to heavy virtual particles 
propagating in an internal loop of the transition. 
Even today, the absence of a Standard Model (SM) tree-level background 
turns precision measurements of meson-antimeson  
mixing into an excellent probe for new physics  
effects; recently, a possible indication of such effects in the phase $\Phi_s$ of  
the $B_{s} - \bar B_{s}$ system \cite{Aaltonen:2007he, Abazov:2008fj} has stirred 
vivid discussions \cite{Charles:2004jd, Lenz:2006hd, Bona:2009tn}. 
 
Among the four mixing systems ($K^0$, $D^0$, $B_{d}$ and $B_{s}$) the $D^0$  
system is in a sense unique. The mixing mechanism relies  
on internal $d$-type quarks; due to the smaller mass of the $b$ quark  
compared to the $t$ quark, the kinematics of the dispersive part of the mixing amplitude  
are not completely dominated by the heavy third generation quark. 
Furthermore, due to the specific structure of the  
Cabibbo-Kobayashi-Maskawa couplings \cite{Kobayashi:1973fv}, 
the absorptive part $\Gamma_{12}$ will feature an extremely efficient  
Glashow-Iliopoulos-Maiani (GIM) mechanism \cite{Glashow:1970gm}. We  
will discuss this in detail later on and show that it leads to a  
suppression of the leading contribution by several orders of magnitude. 
 
On the experimental side, the first evidence for charmed  
meson oscillation was reported by the {\it Belle} collaboration 
\cite{Staric:2007dt, Abe:2007rd}, by {\it BABAR} \cite{Aubert:2007wf} 
and later by CDF \cite{Aaltonen:2007uc}.  
Currently, the relative decay width difference $y$ and the 
relative mass difference $x$ 
have been measured to 
about $20\%$ accuracy. The Heavy Flavor Averaging Group (HFAG) quotes the best-fit values \cite{Barberio:2008fa} 
\begin{equation} 
y := \frac{\Delta \Gamma}{2 \Gamma_{D^0}} = \left(7.3 \pm 1.8\right)\times{10^{-3}}\;, 
\hspace{0.3cm} 
x :=  \frac{\Delta M}{\Gamma_{D^0}} = {9.1^{+2.5}_{-2.6}}\times{10^{-3}}. 
\end{equation} 

The theory status for the $D^0$ system is, unfortunately,  
in a slightly worse shape.  
There are  two main approaches to this issue: Heavy Quark Expansion (HQE) \cite{Shifman:1984wx, Shifman:1986mx, Bigi:1992su} is an expansion of the bilocal  
$\Delta C =2$ matrix elements as a series of local operators of increasing dimension, which are suppressed by powers of the heavy quark mass. 
While this technique provides an excellent tool to study $B$ mixing \cite{Lenz:2008xt}, 
the $D^0$ system predictions \cite{Georgi:1992as, Ohl:1992sr, Bigi:2000wn, Golowich:2006gq} differ from experiment by up to a factor $10^3$. 
There are four main lines of argumentation why such a behavior is observed. 
First of all, a breakdown of the expansion  
in powers of the charm mass, which may not qualify it as ``heavy'' in  
the sense of the expansion, is possible.  
Comparing the ratios of the typical hadronic scale $\Lambda$ and the heavy quark 
mass in the $B$ and $D^0$ system ($\Lambda/m_b \approx 0.05$ versus $\Lambda/m_c \approx 0.25$),  
the expansion parameter has increased by  almost a factor of five.  
Secondly, one can invoke a violation of quark-hadron duality  
due to non-perturbative long distance effects; reliable quark-level predictions for the decay width difference  may consequently serve as a probe of quark-hadron duality in the charm system. 
Furthermore, such a deviation could arise, if the severe GIM cancellations  
present in the leading terms of the OPE are lifted for higher dimensional operators. 
Finally, new physics may enhance the SM result for $D^0-\bar D^0$ mixing.

A second way to access the calculation of the  $D^0-\bar D^0$ decay width  difference $\Delta \Gamma$ is    
 based on exclusive  
 techniques \cite{Falk:2004wg, Falk:2001hx}. 
 In principle one has therefore to determine 
 all contributions to $\Gamma_{D^0}$ and $\Gamma_{\bar{D}^0}$ 
 with high precision, which is clearly beyond our current ability. 
 As a first step to determine the size of $\Delta \Gamma$ 
 within the exclusive approach the authors of [30,31] take 
 only the difference of the phase space of the corresponding 
 final states into account. 
 
In this work we reexamine the HQE for the relative decay width difference $y$.  
We begin with a short introduction of the formalism for the $D^0-\bar D^0$ mixing via box diagrams. 
Sect.~\ref{3} deals with the leading-order (LO) HQE predictions for the mixing matrix element  
$\Gamma_{12}$ in next-to-leading order (NLO) in $\alpha_s$.  
Furthermore, we will show that, contrary to expectation,  
the large cancellations due to GIM mechanism  
can generate a sizeable imaginary part in $\Gamma_{12}$. 
After a brief discussion of possible effects of higher order terms in the HQE, 
a new physics model is presented, which can substantially  
enhance the leading HQE term. We finish with a conclusion.

\section{Mixing formalism} 
The mixing of neutral mesons is described by box diagrams with the absorptive part $\Gamma_{12}$ 
and the dispersive part $M_{12}$. The observable mass and decay rate differences are given 
by ($ \phi := \arg [ - M_{12} / \Gamma_{12}  ]$) 
\begin{eqnarray} 
\left( \Delta M \right)^2 - \frac14 \left( \Delta \Gamma \right)^2 
& = & 
4 |M_{12}|^2 - |\Gamma_{12}|^2 , 
\nonumber \\ 
\Delta M \Delta \Gamma  
& = & 
4 |M_{12}| |\Gamma_{12}| \cos (\phi) \, . 
\end{eqnarray} 
If $|\Gamma_{12}/M_{12}| \ll 1$, as in the case of the $B_s$ system ($ \approx 5 \cdot 10^{-3}$) or if 
$\phi \ll 1$, one gets the famous approximate formulae 
\begin{equation} 
\Delta M = 2 |M_{12}|\, , \hspace{0.5cm} \Delta \Gamma = 2 |\Gamma_{12}| \cos \phi \, .  
\label{approx} 
\end{equation} 
The experimental values for $x$ and $y$ suggest that in the $D^0$ system  
$|\Gamma_{12}/M_{12}| \approx {\cal O} (1)$, the size of the mixing phase $\phi$ will be  
discussed below. 
%
\section{Leading HQE predictions} \label{3} 
\begin{figure}[t] 
\centering 
\includegraphics[width=0.36\textwidth,clip, angle=0]{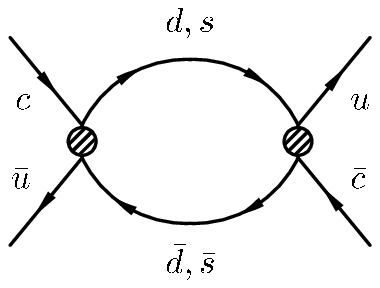} 
\hspace{20mm} 
\includegraphics[width=0.36\textwidth,clip, angle=0]{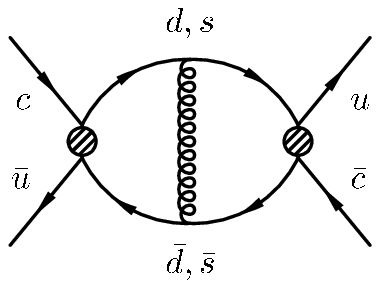} 
\caption{Contributions to $\Gamma_{12}$ from operators of dimension 6 ($D=6$). The leading order  
QCD diagram is shown in the left panel, an example for $\alpha_s$ corrections is shown  
in the right panel.} 
\label{fig:D=6} 
\end{figure}

The absorptive part of the box diagram with internal $s$ and $d$ quarks  
can be decomposed according to the CKM structure as 
\begin{equation} 
\Gamma_{12} =  
- \left( \lambda_s^2 \Gamma_{ss}  
      +2 \lambda_s \lambda_d \Gamma_{sd}  
       + \lambda_d^2 \Gamma_{dd} \right),  
\label{Gamma12} 
\end{equation} 
with $\lambda_x = V_{cx} V_{ux}^*$.  
The application of the heavy quark expansion (HQE), which turned out to be very successful in the  
$B$ system, to the charm system typically meets major doubts.  
Our strategy in this work is the following: instead of trying to clarify the convergence of the HQE 
in the charm system in advance, we simply start with the leading term and determine corrections to it. 
The size of these corrections will give us an estimate for the convergence of the HQE in the $D^0$ system. 
To this end  we first investigate the contribution of dimension-6 ($D=6$) operators 
to $\Gamma_{12}$, see Fig.~\ref{fig:D=6}. 
Next we include NLO-QCD corrections, which were calculated for the $B_s$ system  
\cite{Beneke:1998sy,Beneke:2003az,Ciuchini:2003ww,Lenz:2006hd} and  
subleading terms in the HQE (dimension-7 operators), which were obtained in  
\cite{Beneke:1996gn,Dighe:2001gc}. 
To investigate the size of $\alpha_s$ and $1/m_c$ corrections in more detail we  
decompose $\Gamma_{ss,sd,dd}$ into the Wilson coefficients $G$ and $G_\text{S}$ of the 
$\Delta C=2$ operators $Q$ and $Q_\text{S}$ (for more details see \cite{Lenz:2006hd}) 
\begin{equation} 
\Gamma_{xx}^{D=6,7} = 
 \frac{G_F^2 m_c^2}{24 \pi M_{D^0}} \left[ 
G^{xx} \langle D^0 | Q | \bar{D}^0 \rangle + G_\text{S}^{xx} \langle D^0 | Q_\text{S} | \bar{D}^0 \rangle \right] + \Gamma_{xx}^{\frac{1}{m_c}}, 
\label{oldbasis} 
\end{equation} 
with 
\begin{eqnarray} 
Q   & = & \bar{u}_\alpha \gamma^\mu (1-\gamma_5) c_\alpha \cdot  
          \bar{u}_\beta  \gamma_\mu (1-\gamma_5) c_\beta, 
\nonumber 
\\  
Q_\text{S} & =& \bar{u}_\alpha (1+\gamma_5) c_\alpha \cdot  
         \bar{u}_\beta  (1+\gamma_5) c_\beta. 
\label{DeltaC=2op} 
\end{eqnarray} 
The effect of the QCD corrections has already been discussed in  \cite{Golowich:2005pt}.  
In our numerics we carefully expand in $\alpha_s$: the leading order QCD contribution consists of  
leading order $\Delta C=1$ Wilson coefficients inserted in the left diagram of Fig.~(\ref{fig:D=6}),  
while our NLO result consists of NLO  $\Delta C=1$ Wilson coefficients inserted in both diagrams of  
Fig.~(\ref{fig:D=6}) and consistently throwing away all terms which are explicitly of  
${\cal O}(\alpha_s^2)$. Following \cite{Beneke:2002rj} 
we have also summed terms like $z \ln z$ to all orders; therefore, we use in our numerics 
$\bar{z} = \bar{m}_s(\bar{m}_c)^2 /  \bar{m}_c(\bar{m}_c)^2 \approx 0.0092$. 
The matrix elements in Eq.~(\ref{DeltaC=2op}) are parameterized as 
\begin{eqnarray} 
\langle D^0 | Q          | \bar{D}^0 \rangle & = &   \frac{8}{3} f_{D^0}^2 M_{D^0}^2 B(\mu) \, , 
\\ 
\langle D^0 | Q_\text{S}       | \bar{D}^0 \rangle & = & - \frac{5}{3} f_{D^0}^2 M_{D^0}^2  
                       \left( \frac{M_{D^0}}{m_c(\mu) + m_u(\mu)} \right)^2 B_\text{S}(\mu) \, . 
\nonumber 
\end{eqnarray} 
We take $f_D = 212(14) $ MeV from \cite{Lubicz:2008am} and we derive  
$ B(m_c) = 0.9\pm 0.1$ and $ B_\text{S}(m_c) \simeq 1.3$ from  
\cite{Becirevic,Gorbahn:2009pp}. 
Finally we use the $\overline{\text{MS}}$ scheme for the charm mass, $\bar{m}_c(\bar{m}_c) = 1.27$ GeV.  
For clarity we show in the following only the results for the central values of our parameters, 
the error estimates will be presented at the end of the next section. We obtain 
\begin{displaymath} 
\begin{array}{c|ccc} 
\hline\hline 
              & \mbox{LO} & \mbox{NLO} & \Delta \mbox{NLO} / \mbox{LO} 
\\ 
\hline  
G^{ss}        & \;\;0.25^{+0.09}_{-0.06}\;\; &\;\; 0.37^{+0.18}_{-0.20} \;\; & \;\; + 48 \%\;\; 
\\ 
G^{ds}        & 0.26^{+0.09}_{-0.06} & 0.39^{+0.19}_{-0.21}       &  + 49 \% 
\\ 
G^{dd}        & 0.28^{+0.09}_{-0.06} & 0.42^{+0.19}_{-0.22}       &  + 49 \% 
\\ 
\hline 
G^{ss}_\text{S}     & 1.97^{+0.15}_{-0.29} & 1.34^{+0.19}_{-0.23}       &  - 32 \% 
\\ 
G^{ds}_\text{S}      & 1.98^{+0.15}_{-0.29} & 1.34^{+0.19}_{-0.23}       &  - 32 \% 
\\ 
G^{dd}_\text{S}      & 1.98^{+0.15}_{-0.29} & 1.35^{+0.19}_{-0.23}       &  - 32 \% 
\\ 
\hline\hline 
\end{array} 
\end{displaymath} 
For the error estimate we vary $\mu_1$ between $1$~GeV and $2 m_c$. 
Combining $G$ and $G_\text{S}$ to $\Gamma_{xx}$ we get  
(in units of $\text{ps}^{-1}$) 
\begin{displaymath} 
\begin{array}{c|cccc} 
\hline\hline 
              & \mbox{LO} & \Delta \mbox{NLO-QCD} & \Delta 1/m_c & \sum 
\\ 
\hline  
\Gamma_{ss}        &\;\; 3.52\;\;     & \;\;-0.94  \;  (-27 \%)\;\;   & \;\;-0.76  \;(-22 \%) \;\;  & \;\; 1.82 \;\; 
\\ 
\Gamma_{ds}        & 3.54      &  -0.93  \;(-26 \%)   & -0.76  \;(-22 \%)   & 1.84 
\\ 
\Gamma_{dd}        & 3.55      & -0.92   \;(-26 \%)   & -0.76 \; (-22 \%)   & 1.87 
\\ 
\hline\hline 
\end{array} 
\end{displaymath} 
Using instead the operator basis suggested in \cite{Lenz:2006hd} with 
\begin{eqnarray} 
\tilde Q_\text{S} & =& \bar{u}_\alpha (1+\gamma_5) c_\beta \cdot  
         \bar{u}_\beta  (1+\gamma_5) c_\alpha \, , 
\\ 
\langle D^0 | \tilde Q_\text{S} | \bar{D}^0 \rangle & = &   \frac{1}{3} f_{D^0}^2 M_{D^0}^2   
                       \left( \frac{M_{D^0}}{m_c(\mu) + m_u(\mu)} \right)^2 \tilde B_\text{S}(\mu) \, , 
\nonumber 
\end{eqnarray}  
and $ \tilde B_\text{S}(m_c) \simeq 1.2$  leads to 
\begin{displaymath} 
\begin{array}{c|cccc} 
\hline\hline 
              & \mbox{LO} & \Delta \mbox{NLO-QCD} & \Delta 1/m_c & \sum 
\\ 
\hline  
\Gamma_{ss}        & \;\;1.77\;\;     & \;\;+0.02  \; (+1 \%)\;\;   & \;\;-0.34  \;(-19 \%)\;\;   & \;\;1.46\;\; 
\\ 
\Gamma_{ds}        & 1.78     & +0.03  \; (+2 \%)   & -0.34 \; (-19 \%)   & 1.48 
\\ 
\Gamma_{dd}        & 1.80      & +0.05  \; (+3 \%)   & -0.34 \; (-19 \%)   & 1.51 
\\ 
\hline\hline 
\end{array} 
\end{displaymath} 
All in all we get large QCD (up to $50\%$) and large 1/$m_c$ corrections (up to $30\%$) 
to the leading $D=6$ term, which considerably lower the LO values. 
In the $(Q, \tilde{Q}_S)$-basis numercial cancellations can occur which 
mimic very small QCD corrections for $\Gamma_{xy}$. 
Despite large corrections, the HQE seems not to be completely off. From  
our above investigations we see no hints for a breakdown of OPE.  
The same argument can be obtained  from the comparison of $B$ and $D$ meson lifetimes.  
In the HQE one obtains 
\begin{equation} 
\frac{\tau_B}{\tau_{D^0}} = 
\frac{\Gamma_{0,{D^0}} + \delta \Gamma_{D^0}}{\Gamma_{0,B} + \delta \Gamma_B} 
\approx  \frac{\Gamma_{0,{D^0}}}{\Gamma_{0,B}} \left( 1 + \frac{\delta \Gamma_{D^0}}{\Gamma_{0,D^0}} \right)  
                                     \left( 1 - \frac{\delta \Gamma_B}{\Gamma_{0,B}} \right)\;, 
\end{equation} 
where the leading term $\Gamma_0 \propto m_{b,c}^5 V_{\text{CKM}}^2$ corresponds to the free quark  
decay and all higher terms in the HQE are comprised in $\delta \Gamma$.  
For the ratio $\Gamma_{0,D^0} / \Gamma_{0,B}$ one gets a value close to one.  
Higher order HQE corrections in the $B$ system are known to be smaller than 10 \% \cite{Lenz:2008xt}. 
Using the experimental values for the lifetimes we get 
\begin{eqnarray} 
\frac{\tau_B}{\tau_{D^0}} & \approx &  1.4 ... 4 \, (\mbox{Exp.})  
\approx  1 \cdot \left( 1 + \frac{\delta \Gamma_{D^0}}{\Gamma_{0,D^0}} \right)\;. 
\end{eqnarray} 
From this rough estimate one expects higher order HQE corrections in the 
$D$ system of up to 300 \%. So clearly no precision determination will   
be possible within the HQE, but the estimates should still be within 
the right order of magnitude. 
%
%
%
%
\section{Cancellations}  
As is well known huge GIM cancellations  
\cite{Glashow:1970gm} arise in the leading HQE terms for $D^0$ mixing. To make these effects more obvious,  
we use the unitarity  of the CKM matrix ($\lambda_d + \lambda_s + \lambda_b = 0$) to rewrite the expression 
for the absorptive part in Eq.~(\ref{Gamma12}) as  
\begin{equation} 
\Gamma_{12} = - \lambda_s^2 \left( \Gamma_{ss}\! - 2 \Gamma_{sd} + \Gamma_{dd}  \right) 
+2 \lambda_s \lambda_b \left( \Gamma_{sd} - \Gamma_{dd} \right) 
- \lambda_b^2 \Gamma_{dd} \;. 
\label{cancel} 
\end{equation} 
Note that the CKM structures differ enormously in their numerical values:  
$\lambda_{d,s} \propto \lambda$ and $\lambda_b \propto \lambda^5$  
in terms of the Wolfenstein parameter $\lambda \approx 0.2255$.  
In the limit of exact SU$(3)_\text{F}$ symmetry,  
$\Gamma_{ss} = \Gamma_{sd} = \Gamma_{dd}$ holds and therefore, contrary to many 
statements\footnote{These statements are obtained assuming $V_{ub}$ = 0, 
which we have shown to be not justified for the HQE approach.} 
in the literature, 
$\Gamma_{12} = - \lambda_b^2 \Gamma_{dd}$ is not zero although strongly CKM suppressed. 
%
%
Next we expand the arising terms in $\bar{z}$.  
Using Eq.~(\ref{oldbasis}) we get in LO 
\begin{eqnarray} 
\Gamma_{ss}^{D=6} & = & 3.55477 - 3.22581 \bar{z} - 14.877 \bar{z}^2 + ... \, , 
\nonumber 
\\ 
\Gamma_{sd}^{D=6} & = & 3.55477 - 1.61291 \bar{z} - 7.43849 \bar{z}^2 + ... \, . 
\end{eqnarray} 
The first term in the above equations obviously corresponds to $\Gamma_{dd}^{D=6}$. 
For the combinations in Eq.~(\ref{cancel}) we get 
\begin{eqnarray} 
\left( \Gamma_{ss} - 2 \Gamma_{sd} + \Gamma_{dd} \right)^{D=6} & = & -36.91 \bar{z}^3 \approx \lambda^{7.0}, 
\nonumber 
\\ 
\left( \Gamma_{sd} - \Gamma_{dd} \right)^{D=6} & = &  -1.613 \bar{z} \approx \lambda^{2.8}\,. 
\end{eqnarray} 
To make the comparison with the arising CKM structures more obvious, we have expressed the size of these  
combinations also in terms of powers of the Wolfenstein parameter $\lambda$. 
As is well known, we find in the first term of Eq.~(\ref{cancel}) an extremely effective GIM cancellation, 
only terms of order $\bar z^3$ survive. 
In NLO we get 
\begin{eqnarray} 
\Gamma_{ss}^{D=6,7} & = & 1.8696 - 5.5231 \bar{z} - 13.8143 \bar{z}^2 + ... \, , 
\nonumber 
\\ 
\Gamma_{sd}^{D=6,7} & = & 1.8696 - 2.7616 \bar{z} -   7.4906 \bar{z}^2 + ... \, . 
\end{eqnarray} 
The arising combinations in Eq.~(\ref{cancel}) read now 
\begin{eqnarray} 
\left( \Gamma_{ss} - 2 \Gamma_{sd} + \Gamma_{dd} \right)^{D=6,7}  = & 
1.17 \bar{z}^2 -  59.5 \bar{z}^3 \approx \lambda^{6.2} -  \lambda^{6.7}, 
\nonumber 
\\ 
\left( \Gamma_{sd} - \Gamma_{dd} \right)^{D=6,7} = &  -  2.76 \bar{z} \approx \lambda^{2.5}. 
\end{eqnarray} 
The fact that now the first term of Eq.~(\ref{cancel}) is of order  $\bar z^2$ compared to  
$\bar z^3$ in the case of the LO-QCD value was discussed in detail in \cite{Beneke:2003az} and later on  
confirmed in \cite{Golowich:2005pt}.  
These numbers are now combined with  CKM structures, whose exact values read 
\begin{eqnarray} 
\lambda_d & = &  - c_{12} c_{23} c_{13} s_{12}   - c_{12}^2 c_{13} s_{23} s_{13} e^{i \delta_{13}} 
                 = {\cal O} \left( \lambda^1 + i \lambda^5 \right), 
\nonumber \\ 
\lambda_s & = &  + c_{12} c_{23} c_{13} s_{12} - s_{12}^2 c_{13} s_{23} s_{13} e^{i \delta_{13}}  
                 = {\cal O} \left( \lambda^1 + i \lambda^7 \right), 
\nonumber \\ 
\lambda_b & = &  c_{13} s_{23} s_{13} e^{i \delta_{13}} = {\cal O} \left(\lambda^5 + i \lambda^5 \right), 
\label{CKM3exact} 
\end{eqnarray} 
with $c_{ij} = \cos (\theta_{ij})$ and $s_{ij} = \sin (\theta_{ij})$. 
Looking at Eq.~(\ref{CKM3exact}), it is of course tempting to throw away the small  
imaginary parts of $\lambda_d$ and $\lambda_s$, but we will show below that this is not 
justified. Doing so and keeping only the leading term in the CKM structure 
($c_{12} c_{23} c_{13} s_{12}$), which is equivalent to approximate $\lambda_b = 0$, 
one gets a real $\Gamma_{12}$ which vanishes in the exact SU$(3)_\text{F}$ limit. 
Keeping the exact expressions, we see that the first term in Eq.~(\ref{cancel}) is leading in CKM  
(${\cal O} \left[ \lambda^2 + i \lambda^8 \right]$) and has a negligible imaginary part, but it is suppressed  
by $ 1.2 \bar{z}^2 \approx \lambda^{6.2}$. The second term in Eq.~(\ref{cancel})  
is subleading in CKM  (${\cal O} \left[ \lambda^6 + i \lambda^6 \right]$ 
and it can have a sizeable phase. This term is less suppressed by SU$(3)_\text{F}$ breaking  
($\approx 2.7  \bar z \approx \lambda^{2.5}$).  
The third term in Eq.~(\ref{cancel}) is not suppressed at all by  SU$(3)_\text{F}$ breaking,  
but it is strongly CKM suppressed  (${\cal O} \left[ \lambda^{10} + i \lambda^{10} \right]$).  
For clarity we compare the different contributions of Eq.~(\ref{cancel}) in the  
following table\footnote{With `NLO' we denote the sum of leading order value, $\mathcal{O}(\alpha_s)$ QCD corrections and subleading $\frac{1}{m_c}$ contributions.} 
\begin{displaymath} 
\begin{array}{c|ccc} 
\hline\hline 
 & \;\;\mbox{1$^\text{st}$ term} \;\;  & \;\;\mbox{2$^\text{nd}$ term}\;\;  &\;\; \mbox{3$^\text{rd}$ term}\;\; 
\\ 
\hline  
\mbox{LO}  & 37 \lambda_s^2 \bar z^3  &  3.2 \lambda_s \lambda_b \bar z & 3.6 \lambda_b^2 
\\ 
           & \approx \lambda^{9.0}     & \approx \lambda^{8.4}          & \approx \lambda^{9.1}  
\\ 
\hline 
\mbox{NLO} & 1.17 \lambda_s^2 \bar z^2  &  5.5 \lambda_s \lambda_b \bar z &  1.87 \lambda_b^2 
\\ 
           & \approx \lambda^{8.2}     & \approx \lambda^{8.0}          & \approx \lambda^{9.6}  
\\ 
\hline\hline 
\end{array} 
\end{displaymath} 
From this simple power counting, we see that a priori no contribution to  Eq.~(\ref{cancel}) 
can be neglected. Taking into account the hierarchy of the CKM matrix elements\footnote{ 
Actually $|V_{ub}|$ is numerically of order $\lambda^4$ and therefore  
$\lambda_b \propto \lambda^6$ (see \cite{Bobrowski:2009ng}), but in the literature $V_{ub} = A \lambda^3 (\rho - i \eta)$ 
with small values of $\rho$ and $\eta$ is commonly used.}, we find that the first two terms of  Eq.~(\ref{cancel}) are of similar size, while 
the third term is suppressed. Moreover, the second term can give rise to a large phase  
in $\Gamma_{12}$, while the first term has only a negligible phase. 
To make our arguments more solid we perform the full numerics   
using the CKM values from \cite{CKM-Fitter}  
and obtain for the three contributions  
of  Eq.~(\ref{cancel})  
\begin{align}\label{resultwitherrors} 
10^7 \Gamma_{12}^{D=6,7}  = &  -14.6409 \;+ \;   0.0009 i   && (\mbox{1$^\text{st}$ term}) \nonumber 
\\ 
                            & \;\;\;\;\;\;\!-6.68 \;-\;\;\!\;\;\;  15.8   i && (\mbox{2$^\text{nd}$ term})  \nonumber 
\\ 
                            & \;\;\;\;\;\;\!+0.27\;-\;\;\;\;\!\;  0.28   i  && (\mbox{3$^\text{rd}$ term}) \nonumber 
\\ 
                          = & \;\;-21.1 \;-\;\;\;   16.0   i  &&\hspace{-2cm} = (11 ... 39) \,  e^{-i (0.5... 2.6)} \, . 
\end{align} 
Here we show for the first time the errors; they are estimated by varying $\mu_1$ between $1\;\text{GeV}$ and $2m_c$ and by taking into account the results for both choices of the operator basis.  
The first term in Eq. (\ref{cancel}) turns out to be very sensitive with  
respect to the the exact values of the bag parameters and its real part is  
approximately of the same size as the second term, which features a large imaginary part. 
Furthermore, even the third term can give a non-negligible contribution,  
in particular to the imaginary part.

To summarize, we have demonstrated that the typical approximation $\lambda_b \approx 0$, 
which is equivalent to neglecting the imaginary parts of $\lambda_d$ and $\lambda_s$ is wrong for 
the case of the leading (D=6,7) HQE prediction for $y$ and yields the wrong conclusion that  
$\Gamma_{12}^{D=6,7\;\text{NLO}} $ cannot have a sizeable phase. 
We get for the first terms in the OPE a value for $y$ of  
\begin{eqnarray} 
y^{D=6,7\;\text{NLO}} \leq |\Gamma_{12}| \cdot \tau_{D^0} & = & 4.7 \cdot 10^{-7} \! \! \! ... \, 1.6 \cdot 10^{-6} 
\, . \, \, \, \, \, \,  
\end{eqnarray} 
The range of values  was again estimated on the basis of the renormalization scheme dependence 
and the choice of the operator basis.  
These values are still a factor of  
$0.5 ... 1.6 \cdot 10^4$ smaller than the experimental number. 
This is in contrast to our previous expectations that the HQE should give at least the right  
order of magnitude. Moreover, we do not confirm the observation made in \cite{Golowich:2005pt}  
that the NLO result for $\Gamma_{12}$ is almost an order of magnitude larger than the LO result. 
%
%
\section{Higher HQE predictions} 
\begin{figure}[t] 
\centering 
\includegraphics[width=0.36\textwidth,clip, angle=0]{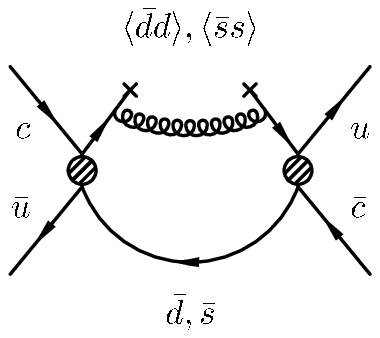} 
\hspace{20mm} 
\includegraphics[width=0.36\textwidth,clip, angle=0]{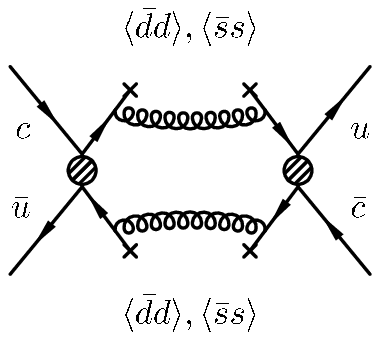} 
\caption{Contributions to $\Gamma_{12}$ from operators of dimension 9 ($D=9$, left panel) and dimension 
12 ($D=12$, right panel). To obtain an imaginary part the $D=9$ diagrams have to be dressed with  
at least one gluon and the $D=12$ diagrams with at least 2 gluons.} 
\label{fig:D=9} 
\end{figure}

In \cite{Georgi:1992as,Ohl:1992sr,Bigi:2000wn} higher order terms in the HQE of $D^0$ mixing  
were discussed. If the GIM cancellation is not as effective as in the leading HQE term,  
operators of dimension 9 and dimension 12, see Fig.~(\ref{fig:D=9}),  might be numerically dominant. 
In order to obtain an imaginary part of the loop integral, the operators of dimension 9  
have to be dressed with at least one gluon and the operators of dimension 12 with at least two gluons. 
If we normalize the leading term (left figure of Fig.~(\ref{fig:D=6})) to 1, we 
expect the $D=9$ diagram of Fig.~(\ref{fig:D=9}) to be of the order 
${\cal O } (\alpha_s (4 \pi) \langle \bar{q} q \rangle   /m_c^3) \approx 0.03$ 
and the  $D=12$ diagram of Fig.~(\ref{fig:D=9}) to be of the order 
${\cal O } (\alpha_s^2 (4 \pi)^2 \langle \bar{q} q \rangle^2   /m_c^6) \approx 10^{-3}$. 
As explained above, the formally leading term of $D=6$ is strongly GIM suppressed to a value of 
about $2 \cdot 10^{-5}$ and the big question is now how severe are the GIM cancellations 
in the $D=9,12$ contribution. For the contributions to $y$ we get the naive expectations 
\begin{displaymath} 
\begin{array}{c|cc} 
\hline\hline 
y    & \;\;\mbox{no GIM} \;\;     &\;\; \mbox{with GIM}\;\; 
\\ 
\hline 
D=6,7  &    2 \cdot 10^{-2} &   5 \cdot 10^{-7} 
\\ 
D=9  &      5 \cdot 10^{-4} &    ? 
\\ 
D=12  &     2 \cdot 10^{-5} &    ? 
\\ 
\hline\hline 
\end{array} 
\end{displaymath} 
If there would be no GIM cancellations in the higher OPE terms, then the $D=9$ or $D=12$ contributions 
could be orders of magnitudes larger than the $D=6$ term, but in order to explain the experimental 
number still an additional  numerical enhancement factor of about 15 has to be present. 
For more substantiated statements $\Gamma_{12}^{D=9,12}$ has to be determined explicitly, 
which is beyond the scope of this work \cite{D=9}. 
This calculation is also necessary in order to clarify to what extent the large phase 
in $\Gamma_{12}$ from the first OPE term will survive. In order to determine the possible SM  
ranges of the physical phase $\phi$, in addition one has to determine $M_{12}$. 
%
%
\section{New physics} 
\begin{figure}[t] 
\centering 
\includegraphics[width=0.4\textwidth,clip, angle=270]{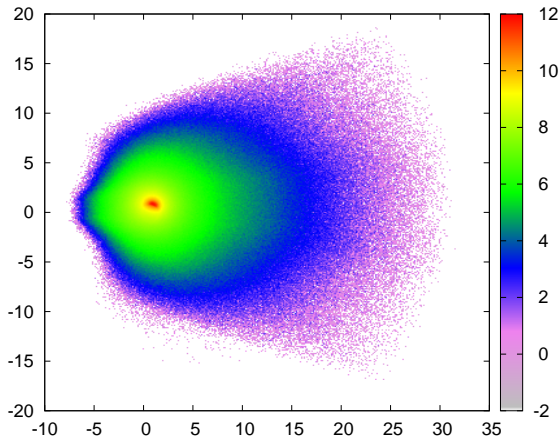} 
\caption{The enhancement factor $\Gamma_{12}^{\text{SM4}}/\Gamma_{12}^{\text{SM3}}$ using the possible values for  
$V_{\text{CKM}4}$ found in \cite{Bobrowski:2009ng}. The color encoded scale denotes the logarithm of the number  
of allowed parameter points of $V_{\text{CKM}4}$.} 
\label{fig:4gen} 
\end{figure}

Finally we would like to address the question, whether new physics (NP) can enhance $\Gamma_{12}$. 
In the $B_s$ system it is argued \cite{Grossman:1996era} that $\Gamma_{12}$ is due to real intermediate 
states, so one cannot have sizeable NP contributions. Moreover, the mixing phase 
in the $B_s$ is close to zero, so the cosine in Eq.~(\ref{approx}) is close to one and therefore 
NP can at most modify $\phi$, which results in lowering the value of $\Delta \Gamma$ compared  
to the SM prediction. In principle there is a loophole in the above argument.  
To $\Gamma_{12}$ also $\Delta B =1$ penguin operators contribute, whose Wilson coefficients  
might be modified by NP effects. But these effects would also change all tree-level $B$ decays. Since this  
is not observed at a significant scale, it is safe to say that within the hadronic uncertainties 
$\Gamma_{12} = \Gamma_{12}^\text{SM}$ and therefore the argument of \cite{Grossman:1996era} holds. 
Since in the $D^0$ system the QCD uncertainties are much larger, also the possible effects might be larger 
but not dramatic.  
The peculiarity of the $D^0$ system -- the leading term in the HQE is strongly suppressed due to GIM  
cancellation -- gives us however a possibility to enhance $\Gamma_{12}$ by a large factor, 
if we manage to soften the GIM cancellation. This might be accommodated either by 
weakening the SU$(3)_\text{F}$ suppression in the first two terms of Eq.~(\ref{cancel}), 
see, e.g., Petrov et al. \cite{Golowich:2006gq}, or by enhancing the CKM factors of the last two terms in 
Eq.~(\ref{cancel}). 
The latter can be realized in a model with an additional  
fourth fermion family (SM4). The usual CKM matrix is replaced by a four dimensional one 
($V_{\text{CKM}4}$) and the unitary  
condition now reads $\lambda_d + \lambda_s + \lambda_b + \lambda_{b'} = 0$. Eq.~(\ref{cancel}) is replaced by 
\begin{eqnarray} 
\Gamma_{12} & = & - \lambda_s^2 \left( \Gamma_{ss}\! - 2 \Gamma_{sd} + \Gamma_{dd}  \right) 
\label{cancel2} 
\\ && 
              + 2 \lambda_s (\lambda_b + {\lambda_{b'}}) \left( \Gamma_{sd} - \Gamma_{dd} \right) 
              - (\lambda_b + \lambda_{b'})^2 \Gamma_{dd} \;. 
\nonumber 
\end{eqnarray} 
In \cite{Bobrowski:2009ng} an exploratory study of the allowed parameter space of $V_{\text{CKM}4}$ was performed 
and as expected only very small modifications of $\lambda_d$ and $\lambda_s$ are experimentally allowed. 
In almost all physical applications these modifications are numerically much smaller than the 
corresponding hadronic uncertainties and therefore invisible. However, in  
the $D^0$ mixing system it might happen, that all dominant contributions cancel and only these modifications survive. 
In the SM the first two terms of Eq.(\ref{cancel}) are numerical equal. In the SM4 the numerical  
hierarchy depends on the possible size of $\lambda_{b'}$, see Eq.(\ref{cancel2}). In particular, it was  
found in \cite{Bobrowski:2009ng} that currently a value of $\lambda_{b'}$ of the  
order $\lambda^3$ is not excluded. 
This means that the second term of Eq.(\ref{cancel2}) could be greatly enhanced by the existence  
of a fourth family and also the third term would now become relevant. 
Using experimentally allowed data points for $V_{\text{CKM}4}$ from \cite{Bobrowski:2009ng}  
we have determined the possible values of $\Gamma_{12}$ in the SM4 from Eq.~(\ref{cancel2}); 
enhancement factors of a few tens are  possible, see Fig.(\ref{fig:4gen}). 
It should be noted that this enhancement is very sensitive to the exact values of the 
bag parameters; e.g.~a reduction of $B_S$ by $30\%$ triples the range of possible enhancement 
factors.  
\section{Discussion and Conclusions} 
In this work we investigated the leading HQE contribution to the absorptive part of 
$D$ mixing and the leading corrections to it. We found that the size of these corrections is large,  
but not dramatic  ($\approx 50\%$ QCD, $\approx 30\% \, \, 1/m_c$). So we see no signal for a  
breakdown of the OPE and it seems that the HQE might be appropriate to estimate the order  
of magnitude of $\Gamma_{12}$. For a further investigation of the question of the convergence of the HQE  
for the case of the $D$ system a systematic study of $D$ meson lifetimes within that framework  
might be very helpful. 
\\ 
Above we have explained in detail that $\Gamma_{12}^{D=6,7}$ gives, due to huge GIM cancellations,  
a value of $y$ which is about a factor of about 10000 smaller than the current experimental  
expectation, but it can have a large phase and it also does not vanish in the  
exact SU$(3)_\text{F}$ limit. The possibility of a sizeable phase in $\Gamma_{12}$  
is a new result. The important question is now, how big can the physical phase $\phi$ be? 
\\ 
Due to the peculiarity of the $D^0$ system---the extreme GIM cancellations---it  
might also be possible that the HQE result is dominated by $D=9$ and $D=12$ contributions,  
if there the GIM cancellations are less pronounced, see e.g. \cite{Bigi:2000wn}. 
To quantify that possibility these higher dimensional corrections have to be determined 
explicitly, i.e beyond the estimates presented in  \cite{Georgi:1992as,Ohl:1992sr,Bigi:2000wn}.
This calculation will also show what values are possible for the phase of $\Gamma_{12}$. 
To determine finally the physical phase $\phi$ one has to determine in addition $M_{12}$. 
 
Currently, estimates about the possible size of the phase in $D$ mixing are typically based on 
the assumption that the $\lambda_b$-term can be neglected. The remaining contribution
proportional to $\lambda_s^2 (\Gamma_{ss} - 2 \Gamma_{sd} + \Gamma_{dd}$ leaves almost no space
for a CP-violating phase. In this paper we have shown that this widely used assumption is wrong 
for the leading (D=6,7) HQE term: the $\lambda_b$-term is as sizeable as the pure $\lambda_s$ term
and it can have a huge phase.
Although we can not  proof at the current stage that a sizeable phase can survive after all 
corrections are included in the HQE calculation, we think one should meet claims that 
already small values for a  $D$ mixing phase are an unambiguous sign for new physics 
with some caution. This question has  to be studied in more detail both within the 
inclusive approach---as explained above---and within the exclusive approach. 
 
To become more concrete, let us speculate: 
if the first term in Eq.~(\ref{cancel}) is enhanced by a factor $\eta^3$ due to higher dimensional  
terms in the HQE, where the GIM-suppression is much weaker compared to the leading HQE terms,  
as advocated e.g. in \cite{Bigi:2000wn},  the same effect---but less pronounced---will also be active  
in  the second term of Eq.~(\ref{cancel}).  
For the first term, the authors of \cite{Bigi:2000wn} see a possibility of enhancing the effect of 
order $z^2$ up to $\sqrt{z}$ (by 3 powers of $\sqrt{z}$). For the second term we expect an enhancement from $z$ to 
$\sqrt{z}$ (by 1 power of $\sqrt{z}$). So the enhancement of the  second term of Eq.~(\ref{cancel}) is a  
factor of the order of $\eta$. 
Now we can make some numerical predictions for $y$ and Im $\Gamma_{12} / $ Re $\Gamma_{12} $ depending on the numerical  
enhancement factor $\eta$ - we use only values for $\eta$ that are within the estimates of  \cite{Bigi:2000wn}: 
\begin{equation} 
\begin{array}{|c|ccc|} 
\hline 
\eta & \frac{\mbox{Im} \Gamma_{12}}{ \mbox{Re} \Gamma_{12}} & y^{\mbox{Theory}}  & \mbox{Deviation from exp. central value of}\;  y  
\\ 
\hline 
\hline 
18.24  & \hspace{0.4cm} 0.32 \%  \hspace{0.4cm} & \hspace{0.4cm} 0.73 \%   \hspace{0.4cm} & \hspace{0.4cm} 0 \, \sigma \hspace{0.4cm} 
\\ 
14.54  & \hspace{0.4cm} 0.50 \%  \hspace{0.4cm} & \hspace{0.4cm} 0.37 \%   \hspace{0.4cm} & \hspace{0.4cm} 2 \, \sigma \hspace{0.4cm} 
\\ 
11.64  & \hspace{0.4cm} 0.78 \%  \hspace{0.4cm} & \hspace{0.4cm} 0.19 \%   \hspace{0.4cm} & \hspace{0.4cm} 3 \, \sigma \hspace{0.4cm} 
\\ 
4.33   & \hspace{0.4cm} 5.62 \%  \hspace{0.4cm} & \hspace{0.4cm} 0.01 \%   \hspace{0.4cm} & \hspace{0.4cm} 4 \, \sigma \hspace{0.4cm} 
\\ 
\hline 
\end{array} 
\label{specu} 
\end{equation} 
We see that the new contribution, we have found in this paper could lead to a relative size of the imaginary part of $\Gamma_{12}$  
of the order of up to $1 \%$, while the absolute value of $\Gamma_{12}$ is close to the experimental value of $y$. 
A size of the physical phase $\phi$ of the order of $1\%$ within the standard model is typically excluded in the literature, 
where the correction we have worked out in this paper is not taken into account. 
But we would like to warn the reader: for the numbers in Eq.~(\ref{specu}) we have purely speculated about the size of  
the enhancement factor $\eta$ - but they  are within the estimates of  \cite{Bigi:2000wn}. Its concrete value has to be 
determined by a calculation of the higher order terms in the HQE. This task, however, is beyond the scope of the current paper; 
we plan to investigate this  question in \cite{D=9}.

Finally we have shown that new physics, in particular a small violation of the unitarity 
of the $3 \times 3$ CKM matrix, can enhance the leading HQE prediction for  
$\Gamma_{12}$ by a  double-digit factor. 
%
\begin{acknowledgments} 
We are grateful to I. Bigi, V. Braun, A. Petrov  and N. Uraltsev for clarifying discussions. M.B. gratefully acknowledges the support by grants of the Freistaat Bayern
(BayBFG) and Studienstiftung des Deutschen Volkes, Bonn/Germany.
J.~Riedl is supported by a grant of the Cusanuswerk, Bonn, Germany.
\end{acknowledgments} 

\providecommand{\href}[2]{#2}

\end{document}